# Electron doping induced stable ferromagnetism in two-dimensional GdI$_3$ monolayer

Rong Guo, Yilv Guo, Yehui Zhang, Xiaoshu Gong, Tingbo Zhang, Xing Yu, Shijun Yuan*, and Jinlan Wang*

Key Laboratory of Quantum Materials and Devices of Ministry of Education，School of Physics, Southeast University, Nanjing 21189, China

Corresponding authors. E-mail: †siesta@seu.edu.cn, ‡jlwang@seu.edu.cn

**Abstract**

As a two-dimensional material with a hollow hexatomic ring structure, Néel-type anti-ferromagnetic (AFM) GdI$_3$ can be used as a theoretical model to study the effect of electron doping. Based on first-principles calculations, we find that the Fermi surface nesting occurs when more than 1/3 electron per Gd is doped, resulting in the failure to obtain a stable ferromagnetic (FM) state. More interestingly, GdI$_3$ with appropriate Mg/Ca doping (1/6 Mg/Ca per Gd) turns to be half-metallic FM state. This AFM-FM transition results from the transfer of doped electrons to the spatially expanded Gd-5$d$ orbital, which leads to the FM coupling of local half-full Gd-4$f$ electrons through 5$d$-4$f$ hybridization. Moreover, the shortened Gd-Gd length is the key to the formation of the stable ferromagnetic coupling. Our method provides new insights into obtaining stable FM materials from AFM materials.

## I. INTRODUCTION

Since the experimental verification of the two-dimensional (2D) ferromagnetic (FM) materials Cr$_2$Ge$_2$Te$_6$[1] and CrI$_3$[2], the study of 2D magnetic materials has become an emerging branch of the 2D family due to its great properties at atomically limit in condensed matter and its potential applications in spintronics devices.[3-18] Although more and more 2D FM materials were discovered, such as Fe$_3$GeTe$_2$,[19] MnP,[20] GdI$_2$,[21] Fe$_2$Ti$_2$O$_9$,[22] and so on,[23-26] 2D intrinsic FM materials are still scarce and far less abundant than expected. Anti-ferromagnetic (AFM) materials have nonzero magnetic exchange coupling, and their magnetic moments are antiparallel below the critical temperature. When the energy level occupation is changed by an external charge or the spacing of the magnetic atoms is changed by strain, it is possible that all or part of the magnetic moment arrangement in AFM materials will be transformed into the parallel coupling, leading the formation of FM or ferrimagnetic ground states.[27] Therefore, it may be a feasible way to acquire FM materials from AFM materials since the latter also have spin exchange interaction.

The magnetic ground state of the $GdI_3$ bulk was found to be AFM[28] with Van der Waals layered structure. The $GdI_3$ monolayer could be exfoliated from its layered bulk,[29] and it is a Mott insulator. An AFM phase transition from Néel-type to stripy-type can occur by charge doping in the $GdI_3$ monolayer.[29, 30] By doping Li or Mg atoms, the Gd-$5d$ orbital is partially occupied in the $GdI_3$ monolayer, inducing ferroelasticity, multiferroicity, and magnetoresistivity in the AFM $(GdI_3)_2Li$ and $(GdI_3)_2Mg$ monolayer. Nevertheless, their magnetic ground states are still stripy-type AFM rather than FM materials, due to the intense lattice deformation caused by the Fermi surface nesting.

The shape of the Fermi surface of a metallic material is determined by the crystal structure and the highest electron occupied energy level. In the stripy-type AFM $(GdI_3)_2Li$ and $(GdI_3)_2Mg$ monolayer, the shorter Gd-Gd pairs tend to be FM coupling, whereas the AFM coupling occurs between longer Gd-Gd pairs, [29] so it is necessary to retain the regular hexagonal structure of $GdI_3$ for stable FM coupling. For this reason, the adjustment of doped electron numbers is the only solution to avoid the Fermi surface nesting.

In this work, we quantitatively analyze the relationship between charge doping ratio and Fermi surface nesting in $GdI_3$ monolayer. It reveals that appropriate divalent cations, such as Mg/Ca doping, can transform the AFM $GdI_3$ monolayer into the FM ground state. Under the appropriate charge doping, the ferromagnetic and stable material $(GdI_3)_6Mg/ (GdI_3)_6Ca$ monolayer was obtained. Due to just 1/3 of the hollow sites of the hexatomic Gd-I rings are occupied by Mg/Ca atoms, the $5d$ electron-bridging $4f$ FM coupling is achieved with less structural deformation. The mechanism of the Gd-Gd FM coupling is to obtain doped electrons in the 5*d* orbitals while avoiding the Peierls phase transition. It is further confirmed that it is a stable FM metal material with 100% spin polarization, even under external stress. Considering this mode of AFM-FM transition by charge doping has a certain universality, our results encourage the scheme to construct new FM 2D materials from AFM materials.

## II. METHODS

We performed spin-polarized density functional theory computations using the Vienna ab initio simulation (VASP) package.[31] The electron-electron interaction was treated self-consistently with a generalized gradient approximation (GGA) using a Perdew-Burke-Ernzerhof (PBE) exchange-correlation functional[32] and a GGA + U strategy to describe the strong Coulomb interaction between the half-filled 4*f*-shells of Gd. The onsite Hubbard parameters U and J were set to 9.2 eV and 1.2 eV on Gd-4*f* orbitals, as used in Ref.[21, 29, 33]. This setting has been proven to be a good description of the structure, electronic properties, and magnetism of $GdI_2$ and $GdI_3$. A vacuum space of at least 15 Å along the out-of-plane direction was employed to ensure that the interactions between periodic images are negligible. The Monkhorst-Pack 9 ×9 × 1 k-point sampling for a primitive cell was used for both geometry optimization and electronic calculations. The lattice constants and atomic coordinates were fully relaxed until the total energy and force converged to $10^{-6}$ eV and to $10^{-2}$ eV/Å, respectively. The valence electron structures of Gd, I, Mg and Ca are, Gd: $4f^75d^16s^2$, I: $4d^{10}5s^25p^5$, Mg: $3s^2$, Ca: $4s^2$, respectively. The *ab initio* molecular dynamics (AIMD) simulations under 400 K and volume (NVT) ensemble were performed with a 3 × 3 × 1 supercell, for which the total simulation time lasts for 16 ps with 2 fs time step. The Monte Carlo simulations based on the 2D Heisenberg model[34] were used to investigate the magnetic stability of the $(GdI_3)_6Mg$ monolayer.

## III. RESULTS AND DISCUSSIONS

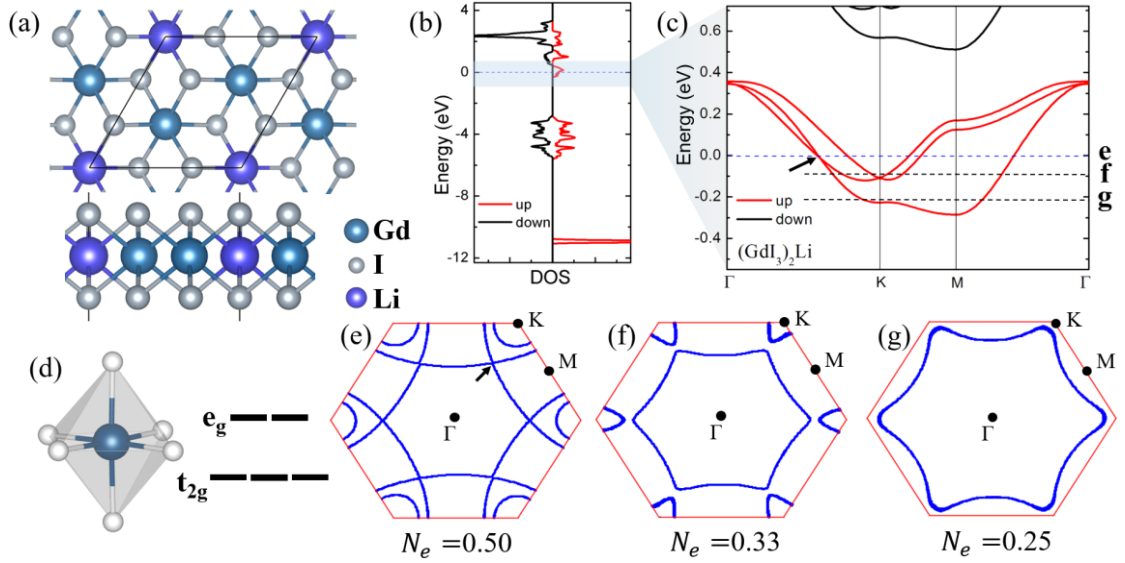

**Fig. 1.** (a) Top and side view, (b) density of states (DOS) and (c) band structure of monolayer $(GdI_3)_2Li$ in FM $P\bar{3}1m$ state. The black arrow indicates the hot spot of Fermi surface nesting. (d) $t_{2g}$ and $e_g$ $d$ orbitals splitting in Gd-$I_6$ octahedron structure. (e)-(g) The 2D Fermi surfaces at the energy of $E_F$, $E_F - 0.11$ eV and $E_F - 0.22$ eV, correspond to the number of doped electrons ($N_e$) of 0.50, 0.33 and 0.25 per Gd atom, respectively.

The $GdI_3$ monolayer consists of hexagonal Gd-I atomic rings. The hollow position of the hexatomic ring provides the space for metal doping. We first consider the structure of Li doping in $GdI_3$. It retains the structure of $P\bar{3}1m$ while limiting $(GdI_3)_2Li$ monolayer to ferromagnetic order, as shown in Fig. 1(a). The Gd atoms are in the center of the octahedron formed by the six I atoms, with the length 3.015 Å for each of six Gd-I bonds. The Gd-$d$ orbitals can be expected to split into double degenerate $e_g$ and triple degenerate $t_{2g}$ in the octahedral crystal field. When Li is incorporated, the $d$ orbitals of $Gd^{3+}$ ion receive electrons, thus the conduction band of $GdI_3$ is partially occupied. Its electronic properties appear as a half-metal, with three spin-up bands crossing the Fermi level, as shown in Fig. 1(c).

In $(GdI_3)_2Li$ monolayer, one $s$ electron of Li transfers into the Gd-$d$ orbital, i.e., each Gd atom gets doped with 0.5 electron. Under this doping condition, the 2D Fermi surface nesting appears [Fig. 1(e)], and six hot spots sit on the line from Γ to K, which is consistent with the previous report.[29] There is no doubt that the Fermi surfaces nesting is caused by multiple bands passing through the Fermi level. Our DFT calculations show that the crossing bands can be reduced to 2 and 1 under the energy level of $E_F - 0.11$ eV and $E_F - 0.22$ eV, corresponding to the doped electron of 0.33 and 0.25 per Gd, respectively [Figs. 1(f) and 1(g)]. By densely sampling the Brillouin zone and counting the number of k points of the occupied states, we can obtain the number of doped electrons (Ne) in the below the specified energy level. Most importantly, the nesting of Fermi surface disappears by shifting the Fermi level below -0.11 eV. It means that if we properly control the number of doped electrons, the Peierls phase transition can be avoided, resulting in a stable FM state $GdI_3$-based material.

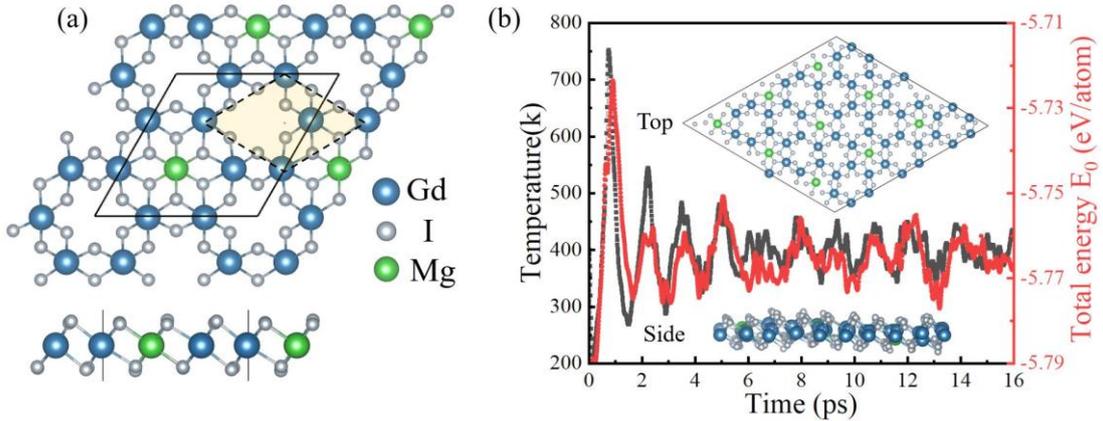

**Fig. 2.** (a) Top view and side view of $(GdI_3)_6Mg$ monolayer. The primitive cell and $\sqrt{3} \times \sqrt{3}$ supercell of $GdI_3$ are indicated by the dashed and solid lines, respectively. (b) Evolution of total energy at 400 K and snapshots of a $(GdI_3)_6Mg$ monolayer after a 16 ps AIMD simulations.

The $(GdI_3)_6Mg$ monolayer is considered with the doping electron 0.33 $e$ per Gd to reduce the perturbation of dopant atoms to the $GdI_3$ hexagonal structure, as shown in Fig. 2(a). The choice is based on the following two considerations: i) Mg has a higher charge doping efficiency, with twice as many valence electrons as Li; ii) One hollow-site dopant atom can transfer its charge to the six nearest Gd atoms in $GdI_3$ monolayer, i.e., the doping ratio can reduce to 1/6. The unit cell of $(GdI_3)_6Mg$ monolayer can be viewed as one Mg atom incorporated into a $\sqrt{3} \times \sqrt{3}$ $GdI_3$ supercell. The hexatomic rings with Mg-inserted expand, while the hexatomic rings without Mg-inserted contract. Due to the low doping concentration, the deformation of hexatomic rings is slight. The different range of the nearest-neighbor Gd-Gd distances in FM state of $(GdI_3)_6Mg$ (3.824 Å ~ 4.527 Å, $\Delta d_{max-min}$= 0.703 Å) is significantly smaller than that of $(GdI_3)_2Mg$ (3.407 Å ~ 4.832 Å, $\Delta d_{max-min}$= 1.425 Å), as listed in Table 1. This means the smaller Mg-induced distortion in FM $(GdI_3)_6Mg$ monolayer.

The stability of a 2D FM crystal is crucial for its experimental fabrication and practical applications. In order to check the stability of the crystal, AIMD simulations were performed to test the thermal stability of the $(GdI_3)_6Mg$ monolayer. As presented in Fig. 2(b), snapshots of the geometries show that the $(GdI_3)_6Mg$ monolayer can keep its original planarity configuration without significant lattice destruction after annealing at 400 K for 16 ps. Small fluctuations of energy with time during the simulations further confirm its thermal stability.

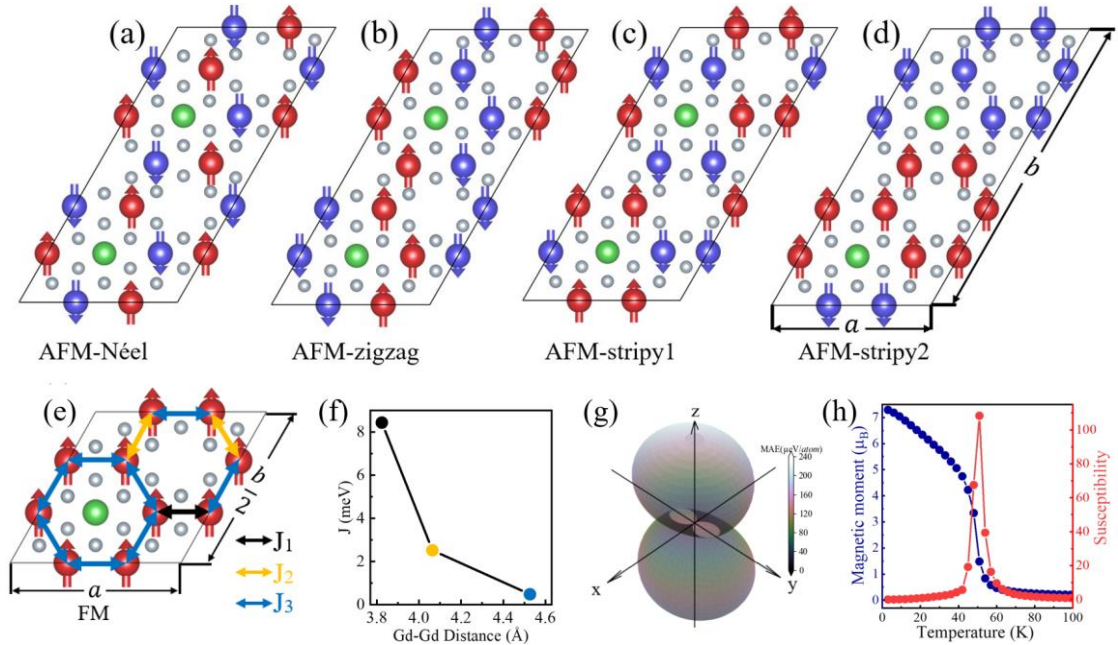

**Fig. 3.** (a-d) Magnetic configurations of Néel-type AFM, zigzag-type AFM, stripy1-type AFM, and stripy2-type AFM $(GdI_3)_6Mg$ monolayer. Red/blue atoms represent Gd atoms with spin-up/down electron configurations, respectively. (e) Magnetic

configuration of FM. The bidirectional arrows correspond to three sorts of Gd-Gd nearest neighbor exchange parameters: $J_1$, $J_2$, and $J_3$. (f) Exchange parameters with different Gd-Gd distances. (g) Angular dependence of the magnetic anisotropy energy (MAE) of the $(GdI_3)_6Mg$. (h) Average magnetic moment per Gd atom (blue) and magnetic susceptibility (red) concerning temperature for $(GdI_3)_6Mg$ monolayer.

**Table 1.** Optimized structures of $(GdI_3)_6Mg$ with different magnetic orders. Lattice constants (a and b) and three sorts of nearest-neighbor Gd-Gd distances ($d_1<d_2<d_3$) are in units of Å. The configuration of $\sqrt{3} \times 2\sqrt{3}$ supercells can be found in Fig. 3(a-e). The energies are in units of meV/f.u., and the FM state is taken as the reference. The lattice parameters of Néel-AFM $GdI_3$ and stripy-type AFM $(GdI_3)_6Mg$ are also listed for comparison.

| Order | a | b | $d_1$ | $d_2$ | $d_3$ | Energy |
|---|---|---|---|---|---|---|
| FM | 13.079 | 26.369 | 3.824 | 4.064 | 4.527 | 0.0 |
| Néel | 13.269 | 26.572 | 4.174 | 4.549 | 4.560 | 345.8 |
| Zigzag | 13.350 | 26.246 | 3.817 | 4.400 | 4.629 | 11.3 |
| Stripy1 | 13.080 | 26.582 | 3.773 | 4.258 | 4.594 | 143.1 |
| Stripy2 | 13.084 | 26.398 | 3.841 | 4.104 | 4.597 | 30.5 |
| $GdI_3$ | 7.785 | 7.785 | 4.494 | - | - | - |
| $(GdI_3)_2Mg$ | 7.787 | 12.416 | 3.407 | 4.832 | - | - |

In order to determine the magnetic ground state of $(GdI_3)_6Mg$ monolayer, four kinds of AFM magnetic ordered states and the FM magnetic ordered state were considered in a $\sqrt{3} \times 2\sqrt{3}$ supercell, as shown in Fig. 3(a)-(e). The structures of FM and AFM $(GdI_3)_6Mg$ monolayer are fully relaxed, with a small difference in lattice constants, in Table 1. The FM state [Fig. 3(e)] possesses the lowest energy. The Néel-type AFM configuration is the most unstable with the Gd-Gd distance in the range 4.17 Å - 4.56 Å, suggesting that the Gd-Gd coupling tends to be FM ordered in this distance range.

Then the coefficient of magnetic exchange interaction was calculated, by comparing the total energies of the different magnetic ordered states in the fixed FM structure. Due to the three different nearest neighbor distances between the magnetic Gd atoms, three exchange parameters are introduced for the $(GdI_3)_6Mg$ monolayer [Fig. 3(e)]. On the basis of the anisotropic Heisenberg model, the spin Hamiltonian is described as

$$H = -J_1 \sum_{\langle i,j \rangle} S_i \cdot S_j - J_2 \sum_{\langle i,k \rangle} S_i \cdot S_k - J_3 \sum_{\langle i,l \rangle} S_i \cdot S_l - \sum_i A(S_i^z)^2$$

Where the first three terms represent the three nearest exchange interactions between the Gd ions, and the last term is the onsite MAE, respectively. The $J_n$ (n=1, 2, 3) is the exchange interaction parameter between sites $i$ and $j$ for three nearest distances. $A$ is the MAE parameter obtained by employing the spin-orbit coupling (SOC) correction. $S_{i,j,k,l}$ is the spin operator.

By mapping the energies in Table 1 and the magnetic moment of each Gd ion (Gd = 7.33 μB) to the Heisenberg Hamiltonian, the coupling parameters of $J_1$ = 8.44 meV, $J_2$ = 2.55 meV and $J_3$ = 0.48 meV, as illustrated in Fig. 3(f). With the increasing of the Gd-Gd distance, the decreasing trend of the J value is remarkable, and the J has a tendency to zero. It is a reasonable extrapolation that the coupling parameter J would turn negative when the Gd-Gd distance is greater than 4.53 Å. This is confirmed by the structural details of the AFM state of $(GdI_3)_6Mg$ and $(GdI_3)_2Mg$. For example, in stripy1-AFM and stripy2-AFM states of $(GdI_3)_6Mg$ monolayer, the longest distance of Gd-Gd nearest neighbors are larger than 4.53 Å ($d_3$ in Table 1), which exactly corresponds to the AFM coupling in these states. In the stripy-AFM $(GdI_3)_2Mg$ monolayer, all the hollow sits of hexagonal Gd-I atomic rings are occupied by Mg atoms, resulting in two sets of significantly different Gd-Gd distances, 3.407 Å and 4.832 Å. The former corresponds to the FM coupling of Gd-Gd pairs, while the latter causes AFM coupling of Gd-Gd pairs for 4.832 Å > 4.53 Å. Therefore, the alternating Gd-Gd FM coupling stripes are formed, and the adjacent stripes are AFM. That is, the ground state of the $(GdI_3)_2Mg$ monolayer is stripy-type AFM.

The doping Mg atoms have two opposite effects. One is the bridging effect. The expanded 5$d$ electrons can couple the magnetic moments of neighboring Gd-4$f$ electrons to form the FM ground state. Another is expansion, that is, the space occupation of Mg atoms expands the hexatomic rings. With the Gd-Gd distance increasing, the coupling between Gd-Gd will change from FM coupling to AFM coupling. For $(GdI_3)_6Mg$ monolayer, 1/3 hexatomic rings expand and 2/3 shrink, dispersing the expansion effect. Meanwhile, every Gd atom has its nearest neighbor doped Mg atom, so the first effect is dominant. When the proportion of doped Mg atoms increases, the latter effect dominates, as in the case of the $(GdI_3)_2Mg$ monolayer.

By evaluating the MAE of the $(GdI_3)_6Mg$ monolayer, the magnetic moment of Gd tends to be along the $x$-axis because of the energy advantage. In Fig. 3(g), the MAE with the magnetic moment lying in the $xy$ plane is significantly lower than that in the $z$-axis direction, about 0.23 meV/Gd. This MAE value is between -0.03 meV/Gd for $GdI_3$ and 1.05 meV/Gd for $(GdI_3)_2Mg$ monolayer.[29] With the inset of Mg atoms, the easy magnetization direction of the $GdI_3$ monolayer changes from out-of-plane to in-plane, and the MAE increases with the increase of the doping content of Mg atoms. Using the exchange parameters and MAE values, a 30×30 supercell is used to simulate the random rotation of spin on each Gd ion. Then the Curie temperature $T_c$ = 51 K of the $(GdI_3)_6Mg$ monolayer is obtained by our Monte Carlo calculation. This temperature is higher than that of the $CrI_3$ monolayer (45 K).[2] The average magnetic moment and susceptibility as a function of temperature are displayed in Fig. 3(h).

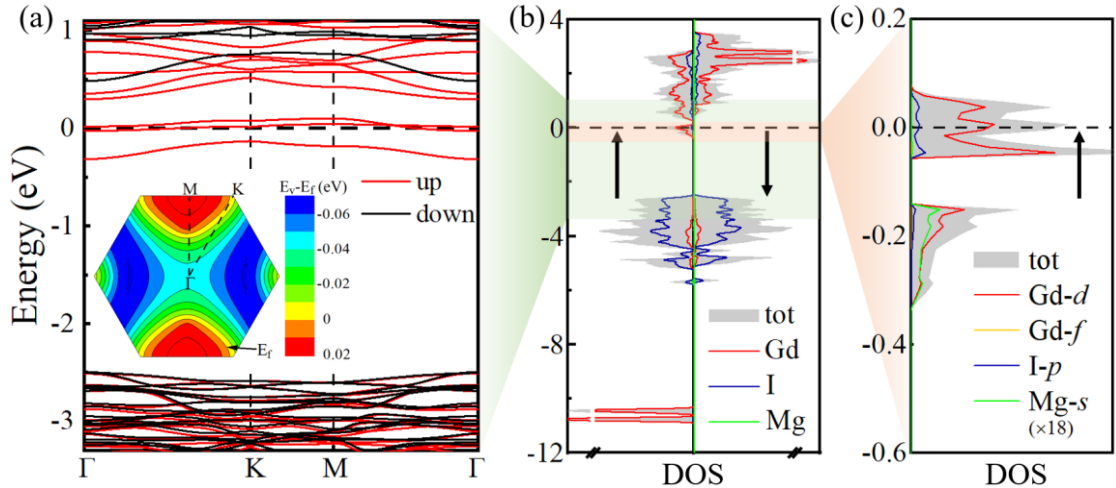

**Fig.4** (a) Electronic band structures of $(GdI_3)_6Mg$ monolayer. The insert shows the energy difference between the valence band $E_v$ and the Fermi level in the 2D Brillouin zone. (b) DOS for the spin majority (↑) and spin minority (↓). The Fermi level is set as zero. (c) Spin-resolved projected DOS around the Fermi level.

The electronic band structures and density of states (DOS) of FM $(GdI_3)_6Mg$ monolayer are calculated, as shown in Fig.4. Clearly, the $(GdI_3)_6Mg$ monolayer is an FM half-metal. It has 100% spin polarization since only the spin-up bands cross the $E_F$, and the spin-down gap is 3.0 eV. Due to the wide excitation gap, the thermal activation effects are difficult to occur, so it may have a high spin injection rate where. By the analysis of the projected density of states (PDOS), the bands across the Fermi level are mainly contributed by the 5*d* orbital of the Gd atoms [Fig. 4(c)]. Due to the Gd-5*d* orbital is localized and less hybridized with *p* orbital of I atoms, the bands near the Fermi level are flat. The 4*f* orbital of the Gd atom is very localized, around the deep level of -11 eV, and the 4*f* electrons provide the dominant magnetic moment of $(GdI_3)_6Mg$. The DOS near the Fermi level also confirms that the magnetism stems from the Gd atom, and the magnetism of other atoms (Mg and I atoms) can be neglected.

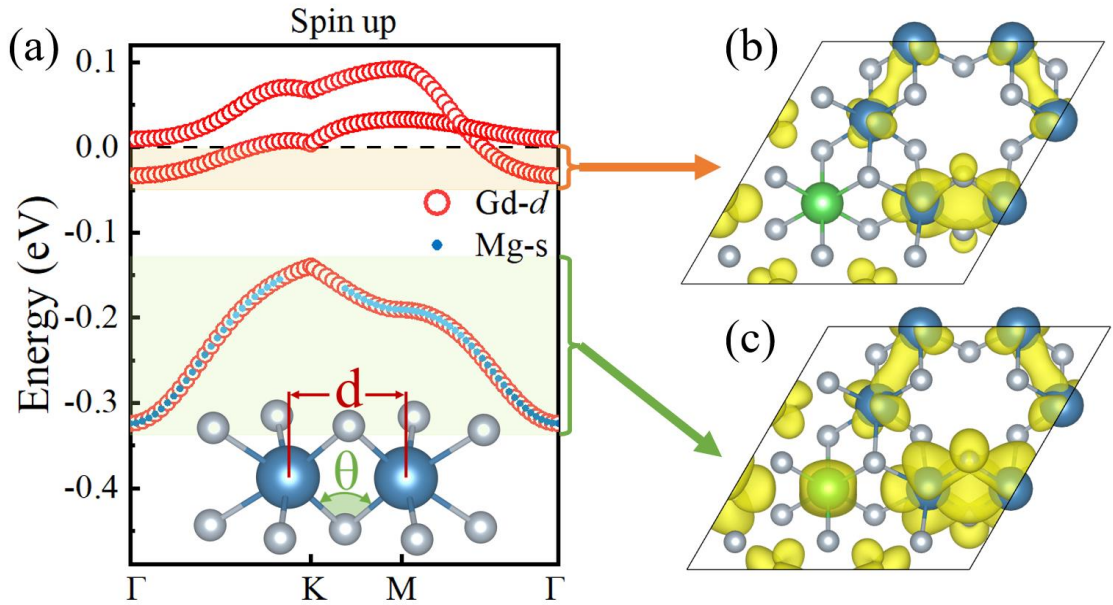

**Fig.5** (a) Three orbital-resolved projected bands with SOC effect near the Fermi level. The inset presents the schematic representation of the distance of Gd-Gd and the angle of Gd-I-Gd. (b, c) The electrons density from -0.05 eV to 0.00 eV (from -0.35 eV to -0.10 eV) of $(GdI_3)_6Mg$ monolayer. The value of the isosurface is 0.0004 $e/Å^3$.

The calculated orbital-resolved projected bands with SOC and electron distribution are shown in Fig.5. The Gd-5$d$ electrons dominate the valence electron structure near the Fermi level. This electron distribution between two adjacent Gd atoms shows the spatial expansion of 5$d$ electrons, while the charge distribution on I atoms is negligible. This further confirms the FM coupling of the magnetic moment of 4$f$ electrons through the nonlocal 5$d$ electrons. The bands in the range of 0.00 eV to -0.05 eV and -0.10 eV to 0.35 eV are derived from the shortest two sets of Gd-Gd pairs, while the longest Gd-Gd pair (4.527 Å) has almost no electron distribution between them. Obviously, the Gd-Gd pairs in the hexagonal Gd-I atomic rings are not equivalent, and this electron distribution is consistent with the exchange parameters J. The densest charge density corresponds to the strongest magnetic coupling.

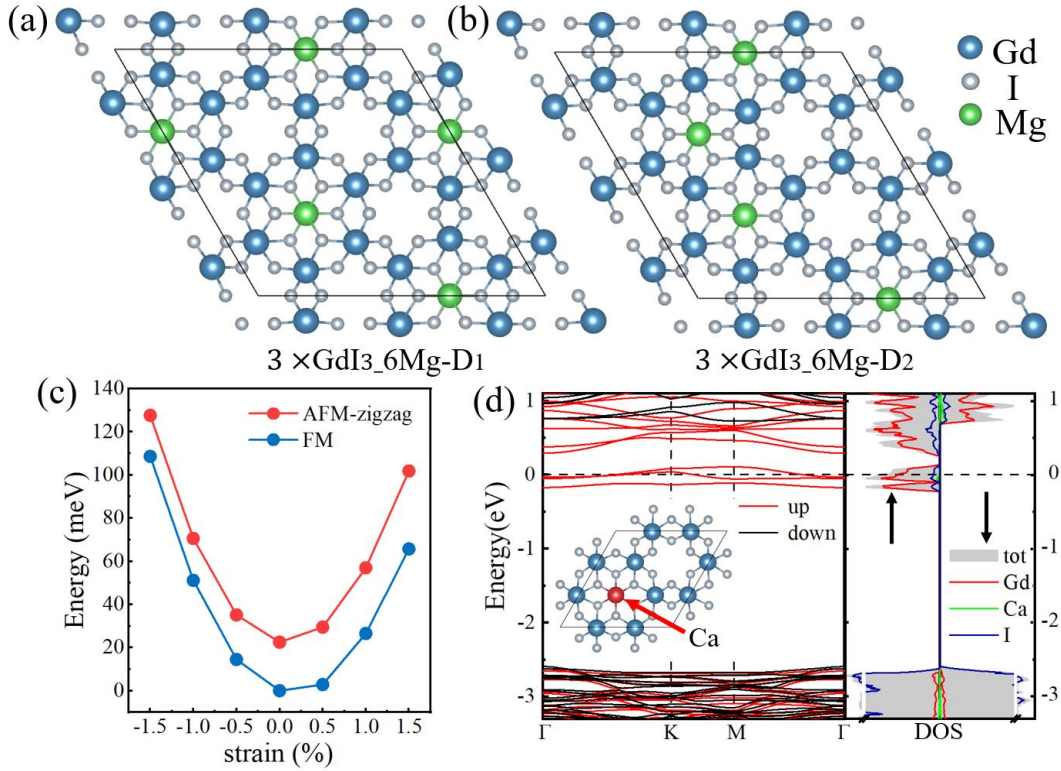

**Fig.6** (a) Doping Mg atoms with a homogeneous configuration and (b) inhomogeneous configuration in 3×3×1 supercell (GdI$_3$)$_6$Mg monolayer. (c) Total energy as a function of biaxial strain for FM and AFM-zigzag (GdI$_3$)$_6$Mg. (d) Band structure and DOS of (GdI$_3$)$_6$Ca monolayer. The inset presents the crystal structure. The Fermi level is set as zero.

The angle of Gd-I-Gd in the pristine $GdI_3$ monolayer is 86.6°, and the Gd-Gd distance is 4.49 Å. In the $(GdI_3)_6Mg$ monolayer, the stronger magnetic exchange interactions correspond to smaller angles and shorter Gd-Gd distances. For $J_1>J_2>J_3$, the Gd-Gd distances are 3.824 Å, 4.064 Å, and 4.527 Å, with the Gd-I-Gd angles of 78.29°, 84.11°, and 95.42°, respectively. This is also confirmed by the electron density distribution [Fig. 5(b) and 5(c)]. The electrons near the Fermi level mainly come from the Gd-Gd pair corresponding to $J_1$ and $J_2$.

To investigate the stable FM $(GdI_3)_6Mg$ configuration, various structures of Mg doped in $3 \times 3 \times 1$ $GdI_3$ supercell were calculated. We found that all the disordered configurations have higher total energy than the ordered configuration. One of disordered configurations is shown in Fig. 6(b), and its total energy is 234 meV higher than that of the homogeneous configuration (Fig. 6(a)). This indicates that the doped Mg atoms tend to be uniformly distributed in the $GdI_3$ monolayer. Furthermore, 2D materials often have structural deformation due to the interaction with substrates, so the energies of the FM and AFM-zigzag $(GdI_3)_6Mg$ under the biaxial strain are calculated. The FM state remains the ground state both under tension and compression conditions [Fig. 6(c)]. These results indicate that the FM $(GdI_3)_6Mg$ structure may be easily observed in experiments and its FM properties are robust. In addition, we also considered the $GdI_3$ monolayer doping with other divalent cations, such as $Ca^{2+}$ [in Fig. 6(d)]. Then the Curie temperature $T_c$ = 12 K of the $(GdI_3)_6Ca$ monolayer is obtained by our Monte Carlo calculation. The $(GdI_3)_6Ca$ monolayer presents a similar band structure to the $(GdI_3)_6Mg$ monolayer, and the FM half-metals state is also more energetically stable than various AFM states. Comparing the doping of Mg with Ca in $GdI_3$ monolayer, the results show that the Mg-doping is superior to the Ca-doping, because $(GdI_3)_6Mg$ can preserve ferromagnetic order at higher temperatures. Due to the ionic radius of $Mg^{2+}$ is smaller than that of $Ca^{2+}$, less lattice distortion occurs on the hexatomic ring structure of $GdI_3$ monolayer in the Mg-doping case. It demonstrates that it is feasible to realize FM 2D materials by electron doping in $GdI_3$ monolayer.

## IV. CONCLUSION

In summary, the present theoretical studies on the 2D GdI$_3$ monolayer reveal that an appropriate amount of electron doping like divalent cations Mg and Ca can convert AFM semiconductors GdI$_3$ monolayer into FM. The key to this transition is to not only partially fill the Gd-5$d$ orbitals but also avoid excessive deformation of the hexatomic ring of the GdI$_3$ lattice. The ratio of one doped Mg/Ca atom for every six GdI$_3$ units achieves exactly this equilibrium. In this case, all Gd-Gd couplings in the (GdI$_3$)$_6$Mg/(GdI$_3$)$_6$Ca monolayer are FM, which stem from the 4$f$-5$d$-4$f$ exchange interaction, and its calculated Curie temperature of (GdI$_3$)$_6$Mg can be as high as 51K. In addition, the homogeneous Mg distribution configuration was confirmed to be stable under strain. These results indicate the sensitive effect of electron doping on the coupling of structure in 2D materials, and this method also can be used to fabricate stable 2D FM materials experimentally.

## ACKNOWLEDGMENTS

This work is supported by the National Key Research and Development Program of China (2022YFB3807203), Natural Science Foundation of China (22033002, 21973011). The authors thank the computational resources from the Big Data Center of Southeast University and National Supercomputing Center of Tianjin. S. Yuan thanks S. Dong for useful discussions.